\documentclass[12pt]{article}
\usepackage{graphicx}
\usepackage[latin1]{inputenc}
\pdfoutput=1
\usepackage[english]{babel}
\usepackage{amsmath}	
\usepackage{amssymb}	

\def\be{\begin{equation}}
\def\ee{\end{equation}}

\def\beq{\begin{equation}}
\def\eeq{\end{equation}}

\begin{document}


\vspace*{-1cm}
\thispagestyle{empty}
\begin{flushright}
\end{flushright}

\begin{center}
{\Large 
{\bf Fusion of conformal interfaces}}
\vspace{2.0cm}

{\large C.~Bachas${}^{\, \sharp}$}%
\hspace*{0.3cm} {\large and} \hspace*{0.2cm} {\large I.~Brunner${}^{\, \flat}$}%
\vspace*{0.5cm}

${}^\sharp$ Laboratoire de Physique Th\'eorique de l'Ecole 
Normale Sup\'erieure 
\footnote{Unit\'e mixte de recherche (UMR 8549)
du CNRS  et de l'ENS, associ\'ee \`a l'Universit\'e  Pierre et Marie Curie et aux 
f\'ed\'erations de recherche 
FR684  et FR2687.}
\\
 24 rue Lhomond, 75231 Paris cedex, France\\
\vspace*{0.4cm}

${}^\flat$ Institut f\"ur Theoretische Physik, ETH-H\"onggerberg\\
8093 Z\"urich, Switzerland\\
\vspace*{2cm}

{\bf Abstract}
\end{center}

We study the fusion of conformal interfaces
in the $c=1$ conformal field theory. 
We uncover an elegant structure reminiscent of that of black holes in
supersymmetric  theories. The role of the BPS black holes  is played by topological interfaces, which
(a) minimize the entropy function, (b) fix  through  an attractor mechanism  one or both
of the bulk radii, and (c) are (marginally) stable under splitting. One significant
difference is that the conserved charges are logarithms of natural numbers, rather
than vectors in a charge lattice, as for BPS states. 
 Besides potential applications
to condensed-matter physics and number theory, 
these  results point to the existence of large solution-generating
 algebras in  string theory.

\newpage

\section{Introduction}

   Conformal interfaces in two dimensions \cite{bbdo}  
are  scale invariant junctions of two conformal field theories.  
They are  generalizations of
conformal defects  and of conformal boundaries  which correspond, respectively,
to the  case of identical theories on the two sides, 
\footnote{In the literature,  general
 interfaces are sometimes also referred to as defects. We believe it is useful  to
 distinguish the two, and not only for semantic reasons.  Defects live at a given point
 in the moduli space of CFTs, and can be always multiplied together. 
 General interfaces,  on the other hand, are  intertwiners between different CFTs.}
or of a  trivial
theory   (with no massless degrees of freedom) on one side.   
  There is an extensive literature,  and many beautiful experimental realizations of 
  such  objects in
 condensed-matter physics  (for reviews and many references see 
for instance \cite{Saleur,
   Affleck}). Applications to condensed-matter physics are outside our
 scope in this work.  
\vskip 1mm 
  
  Two or more interfaces between 
 the same pair of theories  can be added. This amounts to endowing them with
 a finite-dimensional space of (Chan-Patton or ``quantum-dot'') degrees of freedom.    
 Furthermore, an interface between CFT$_1$ and CFT$_2$, 
  and an interface between CFT$_2$ and CFT$_3$
 can, in principle,  be  fused 
  to produce  a CFT$_1$$\to$CFT$_3$ interface. 
   The process is in
general singular, because fusion  (or its inverse,  dissociation) corresponds to
non-trivial renormalization-group flow.   An exception to this rule occurs when one
of the interfaces transmits all incident energy, in which case the left- and right-
Virasoro charges are separately conserved.
Interfaces  of this type,  first introduced
by Petkova and Zuber \cite{Petkova},  can move  freely on a Riemann surface 
and are,  in this sense,  "topological". Their fusion is  non-singular. 
 Many examples of  conformal and topological interfaces have been
 worked out in the literature over the past few years
 (a  list of references is  [5-18]). Lifts to topological
gauge  theories in higher dimensions  [19-22], and
   dual holographic interpretations    
  [1, 23-27]  have been also analyzed.  
 \footnote{For an entry into the extensive
  literature on superconformal defects and the AdS/CFT 
 correspondence in $d=4$  we refer the reader to the review \cite{Kirsch}.}  

\vskip 1mm 
 
   One of the most interesting aspects of
    topological interfaces  is the fact that they are universal maps
 transforming one set of D-branes into another  \cite{Graham, BG}.  All the symmetry
  transformations
 of a CFT can be, in particular, implemented in this fashion  \cite{juerg}. 
 A generic topological interface does not, however, correspond always to 
 a symmetry:  its action  changes the mass, charges and 
 other properties of  the D-branes,  and possibly even those of the  bulk geometry. 
This makes it tempting to speculate
 \cite{GGI}  that the  algebra of all conformal interfaces is  a 
 solution-generating algebra of string theory,   similar  to the
 Ehlers-Geroch transformations  of General Relativity.  
A  classical-geometric intepretation for this algebra
 has been suggested  in ref. \cite{BG}. 
It is based on the folding trick  \cite{AO,bbdo},  
which identifies an interface  with  a middle-dimensional brane  in  the product
target space ${M}_1\times {M}_2$.  
Such a brane can be described, at least locally, in terms of a multiple embedding  of  
  $M_2$  into $M_1$.  \footnote{Assuming for simplicity that the world-volume gauge fields
  are zero.} This embedding determines  the image of $M_2$,   
  and  of  all its D-brane submanifolds,  
  under the  interface map.   
   
 \vskip 1mm  
   
   A crucial question  is  whether this story survives  
   quantization,  and in particular the singularities of interface fusion. In this paper
   we will answer the question  in the simplest setting, 
   that of the $c=1$ conformal field theory.
   The boundary states of this model are classified \cite{GabRec}, 
   its  topological interfaces have been studied \cite{mat2},  
and most  calculations can be done explicitly.   Our analysis will confirm the existence of
a conformal-interface algebra, 
 and its geometric interpretation in the classical limit. At the same time, a
   beautiful and unexpected picture
will emerge:  the topological interfaces of this simple model behave in many  ways
like BPS black holes!    They are minima of an entropy function,    
 they freeze by an attractor mechanism 
  \cite{attractor} one or both of the bulk radii,  and they
  are stable against decay to more elementary interfaces. Their algebra is reminiscent
 of the Harvey-Moore algebra of  BPS states \cite{HaMo}. There is, however, one
 significant difference: the conserved  charges of these topological interfaces 
 do not take values in a regular lattice, but they are instead the logarithms of integers. 
  A  quantum gas of  such particles had been imagined  in the past by Julia
 \cite{bernard} in an effort to rephrase the Riemann hypothesis as a problem
 in  statistical mechanics. 
 
 \vskip 1mm 

 Supersymmetry plays no role in our discussion here.   A different  line of 
 approach,  that avoids the problem of singularities,  has been to study the fusion
 of defect lines  in  theories with extended supersymmetry by twisting to a
topological theory, see \cite{KW,saulina} for results on  
$N=4$ gauge theories in four dimensions,
and \cite{ilka} for  $N=2$ theories in two dimensions.  

\vskip 1mm

The structure of our paper is as follows: 
In section 2  we define our conventions,  and review the 
boundary states for    
toroidally-compactified free-boson CFTs. 
In section 3 we describe the unfolding of the $U(1)^2$ symmetric 
boundary states  of the two-scalar theory
 to  intertwining  operators acting on the moduli space of circle compactifications.
 We explain the special role of topological
interfaces,  and point out the analogy with BPS black holes. 
Sections 4 and 5   contain our main results.  
We show there that the fusion of two  symmetric interfaces is well-defined,
and that it  does not depend on the radius of the collapsed region. 
This reduces the calculation of the algebra to the topological case, 
  studied in ref. \cite{mat2}.  
We explain why  the integer interface charges are multiplicatively conserved,  
and discuss interface stability in a way  reminiscent again
 of black holes.  Finally, in section 6  we extend the discussion to topological
 interfaces for which all  CFT moduli are completely fixed, and which have no
 semiclassical limit. The operator that interpolates between the circle
 and orbifold branches is of this type.   A detailed analysis of the extended
 $c=1$ interface algebra is  postponed to future work.


\section{Boundary states of  toroidal  CFT}

\subsection{Dirichlet and Neumann states}

 We will use the boundary-state formalism \cite{bs,bs1}  in which boundary conditions are described
by states in the Hilbert space of the bulk CFT.  Let us start by recalling the expressions of
the boundary states for  a free scalar field compactified on a circle
of radius $R$. The mode expansion of the field  on the cylinder, 
parametrized by $\sigma \in [0,2\pi)$ and $\tau$,  is given by 
\begin{equation}\label{phi}
\phi(\tau,\sigma)  =  \hat \phi_0 + {\hat N\over 2R}\tau+   \hat MR\sigma + \sum_{n=1}^\infty
{i\over 2\sqrt{n}} \, \left( a_n e^{-in(\tau+\sigma)} + \tilde a_n e^{-in (\tau - \sigma)} -  h.c. \right)\ , 
\end{equation}
where $\hat N$,  $\hat M$ are the integer-valued
momentum and winding operators, and $h.c.$ denotes the hermitean conjugate. 
The canonical commutation relations imply
\begin{equation}
[a_n, a^\dagger_m]  =  [\tilde a_n, \tilde a^\dagger_m]  = \delta_{n,m}\ \ \ {\rm and}
\ \ \ [\hat\phi_0 , {\hat N\over R} ] = i\ ,  
\end{equation}
while  the  Hamiltonian reads
\begin{equation}
H = L_0 + \tilde L_0 =   {\hat N^2\over 4R^2}  + \hat M^2 R^2 + \sum_{n=1}^\infty
n ( a_n^\dagger a_n + \tilde a_n^\dagger \tilde a_n) - {1\over 12}\ . 
\end{equation}

The two simplest  boundary states of this theory\footnote{The free-boson theory contains
also boundary states that break all  $U(1)$ symmetries of the bulk  \cite{GabRec}.
We will discuss these in section 6.}
 correspond to 
 the Dirichlet and Neumann boundary 
conditions for $\phi$. They are given by 
 \begin{equation}\label{dstate}
\underline{\rm Dirichlet}:\hskip 1cm 
\vert\hskip -0.5mm \vert {\rm D}0\,  \rangle \hskip-0.6mm \rangle  \ =\   
\prod_{n=1}^\infty   {\rm exp}( a^\dagger_{n}\tilde a^\dagger_{n}) \ 
\Bigl(  {1\over\sqrt{2R}} \sum_{N=-\infty}^\infty  e^{-i{N\over R} \phi_0}  \vert N, 0 \rangle \Bigr)
\end{equation}
\begin{equation}\label{nstate}
\underline{\rm Neumann}:\hskip 0.8cm 
\vert\hskip -0.5mm\vert {\rm D}1 \rangle \hskip-0.6mm \rangle  \ =\   
\prod_{n=1}^\infty   {\rm exp}( - a^\dagger_{n}\tilde a^\dagger_{n}) \ 
\Bigl(  \sqrt{R} \sum_{M=-\infty}^\infty  e^{iM \tilde\phi_0}  \vert 0, M  \rangle \Bigr)
\end{equation}
where $\vert N,M\rangle$ is the normalized ground state in a given momentum and winding sector. 
Using the commutation relations one verifies easily  that
\begin{equation}\label{bcn}
\phi(0,\sigma) \vert\hskip -0.5mm \vert {\rm D}0\,  \rangle \hskip-0.6mm \rangle  \ =\   
\phi_0 \vert\hskip -0.5mm \vert {\rm D}0\,  \rangle \hskip-0.6mm \rangle \ \ \ \ 
{\rm and}\ \ \ 
\partial_\tau\phi(0,\sigma)   \vert\hskip -0.5mm\vert {\rm D}1 \rangle \hskip-0.6mm \rangle  \ =\ 0\ 
\ \ \ \ \forall \sigma\ ,  
\end{equation}
 as claimed. 
The arbitrary parameters $\phi_0$ and $\tilde\phi_0$ are, respectively, the position of the D$0$ brane,
and the dual Wilson line on the D$1$ brane
 (normalized so as to be periodic under $2\pi$ shifts).

The boundary conditions, eq. (\ref{bcn}),  do not determine the normalization of the 
corresponding states.  
This is usually fixed by Cardy's condition \cite{Cardy}, i.e. by the requirement that the
annulus diagram be equal to the finite-temperature partition function in the transverse channel. 
Although the result for the case at hand  is known, it will be useful to work it out 
explicitly. 
Considering  for instance the D$0$  brane, we may  factorize the annulus diagram as follows:
 \begin{equation}
A_{{\rm DD}} \equiv   \langle \hskip-0.6mm \langle {\rm D}0 \vert\hskip -0.5mm \vert\, 
  q^H \, \vert\hskip -0.5mm \vert {\rm D}0\,  \rangle \hskip-0.6mm \rangle \,
  =\, q^{-{1\over 12}}\,  \langle\phi_0\vert q^H\vert\phi_0\rangle \, \prod_{n=1}^\infty 
  \langle 0\vert  e^{q^{2n} a\tilde a} e^{ a^\dagger\tilde a^\dagger}    \vert 0\rangle\ ,
\end{equation}
where  $q=e^{-T}$, 
$\vert\phi_0\rangle$ is  the state
within the parentheses in equation (\ref{dstate}),  i.e. the 
ground state for fixed value of $\hat \phi_0$,  the
$a$ and $\tilde a$ are canonically
normalized lowering operators of a double harmonic-oscillator system  (the same for all values
of $n$),  and  $\vert 0\rangle$ is the harmonic-oscillator ground state. 
To calculate the individual matrix elements we use  the operator identities 
\begin{equation}\label{gauss}
e^{AB} = \int {d^2z\over\pi}\, e^{-z\bar z - zA - \bar z B}\  \ \ \ {\rm if}\ \ \ [A,B]=0 \ ,  
\end{equation}
 and 
$e^A e^B = e^B e^A e^{[A,B]}$  if  $[A,B]$  is a c-number. A simple calculation then gives:
\begin{equation}
  \langle 0\vert  e^{q^{2n} a\tilde a} e^{ a^\dagger\tilde a^\dagger}    \vert 0\rangle\, 
  =\,  \int {d^2z d^2w \over\pi^2 }\, e^{-z\bar z -w\bar w} 
  \langle 0\vert  e^{- zq^na - \bar z q^n\tilde a} e^{-w a^\dagger -\bar w \tilde a^\dagger} \vert 0\rangle
  \nonumber
\end{equation}
\begin{equation} 
=  \,  \int {d^2z d^2w \over\pi^2 }\, e^{-z\bar z -w\bar w}  e^{q^n(zw + \bar z\bar w)} \,  = \, 
{1\over 1-q^{2n}}\ .
\end{equation}
Computing the remaining matrix element, and combining everything leads to
 \begin{equation}
 A_{{\rm DD}}
   =    \, \Bigl( {1\over 2R}  \sum_{N=-\infty}^\infty 
    q^{N^2\over 4R^2}\Bigr)\,   {1\over \eta(q^2)} \ = \ 
   \Bigl( \sum_{M=-\infty}^\infty
    \widetilde q^{\ 4M^2 R^2}\Bigr)\,  {1\over \eta ({\widetilde q}^{\ 2})} \   ,
\end{equation}
where  $\eta$ is the Dedekind  function and $\widetilde q= e^{-\pi^2/T}$. 
The second equality follows from  the modular  properties of
$\eta$ and the Poisson resummation formula.  
The final expression  is a partition function with integer non-negative multiplicities,  
and a unique lowest-energy state. This shows  that  the D0 boundary state has 
been normalized consistently, and  that it  describes an elementary brane.
  \vskip 3mm

\subsection{Branes in the two-scalar model}
 Let us next consider two scalar fields, $\phi^1$ and $\phi^2$,  compactified on two circles 
with radii $R_1$ and $R_2$. Taking the tensor product of the states
(\ref{dstate}) and (\ref{nstate})   gives the four  {\it factorizable} 
 branes   of the theory, which  correspond to independent Neumann or Dirichlet conditions
for each scalar.  Put differently, these describe a D0,  a D2 and two D1 branes,
with the latter wrapping the two elementary cycles of
 the $(\phi^1,\phi^2)$ torus.
The most  general elementary D1-brane winds $(k_1,  k_2)$ times around these cycles, 
where $k_1$ and $k_2$ may be assumed  relatively prime
and $k_1$  positive.  The corresponding  boundary
states can be constructed easily starting with the factorizable $(1,0)$ brane and then 
rotating  by an angle
 \begin{equation}
 \vartheta = {\rm tan}^{-1} ({k_2R_2\over k_1R_1})\ . 
\end{equation}
The result reads
 \begin{equation}\label{k1k2}
 \vert\hskip -0.5mm \vert \, {\rm D}1,  \vartheta \rangle \hskip-0.6mm \rangle  \ =\   
\prod_{n=1}^\infty   (e^{S^{(+)}_{ij} a_n^i \widetilde a_n^j})^\dagger \ 
\Bigl(  {g}^{(+)} \hskip -3mm \sum_{N,M =-\infty}^\infty  
e^{i{N} \alpha -iM \beta}  \vert k_2N, k_1M
\rangle \otimes \vert
-k_1N, k_2M  \rangle \Bigr)
\end{equation}
\vskip 1mm \noindent where $\alpha$ and $\beta$ are position and Wilson-line moduli,
 the ground states in the tensor product correspond  to  $\phi^1$ and $\phi^2$, 
 in this order,   and 
  \begin{equation}\label{Sold}
 S^{(+)}  =  {\cal R}^T(\vartheta)
  \left(\hskip -1mm \begin{array}{cc}  -1 & 0 \\
                              0 & 1  \end{array}\hskip -1mm \right) \hskip -0.6mm
                               {\cal R}(\vartheta)
                               =
                                \left(\hskip -1mm \begin{array}{cc}
  -{\rm cos\, 2\vartheta} &  -{\rm sin\, 2\vartheta} \\
                               - {\rm sin\, 2\vartheta} & {\rm cos\,
                                 2\vartheta}
  \end{array} \hskip -1mm\right)  \ ,                            
 \end{equation}
 where ${\cal R}(\vartheta)$ is the rotation matrix for angle $\vartheta$.
Finally  the normalization constant is      
the $g$-factor  \cite{g} of the boundary state, 
\be\label{gplus}
{g}^{(+)}  \, = \,  \sqrt{k_1^2R_1^2+k_2^2 
R_2^2\over  2R_1R_2}\, =\, {\ell\over \sqrt{2V}}\,  = \,   \sqrt{
 k_1 k_2  \over {\rm sin} 2\vartheta} 
\ee
with  $\ell$  the length of the D1-brane and  $V$
the volume of the torus. The last  rewriting of the $g$-factor, which will be the most useful to us
in the sequel, follows from simple trigonometric identities.   
The reader can easily
 verify that when  $k_2=\vartheta =0$, the state  (\ref{k1k2}) reduces to the tensor product of   
 (\ref{dstate}) with (\ref{nstate}).  The subscript  ``plus''
 refers to the sign of $-{\rm det} S^{(+)}$,  
 or equivalently to minus the parity of the brane dimension.  The  relevance of this   
 (seemingly upside-down) notation  will become obvious  in the following sections.
\vskip 1mm

The other symmetric stable branes of the $c=2$  theory can be obtained from the above D1
branes by a T-duality transformation of one of the scalars. 
With our conventions, the action of this transformation is \footnote{For a 
general discussion  
of the $O(d,d,Z)$ transformations of D-branes  see \cite{OP}.}
\be
R \to {1\over 2R}\ , \ \ \  \tilde a_n \to - \tilde a_n\ ,\ \ \ 
{\rm and}\ \ \  (N, M) \to (M, N)\ .
\ee
T-dualizing one of the fields, say $\phi^1$,  maps  
$\vert\hskip -0.5mm \vert {\rm D}1,  \vartheta \rangle \hskip-0.6mm \rangle$ to the boundary state
 \begin{equation}\label{d0d2}
 \vert\hskip -0.5mm \vert {\rm D}2/{\rm D}0, \theta\,  \rangle \hskip-0.6mm \rangle  \ =\   
\prod_{n=1}^\infty   (e^{S^{(-)}_{ij} a_n^i \widetilde a_n^j})^\dagger \ 
\Bigl(  {g}^{(-)}\hskip -3mm  \sum_{N,M =-\infty}^\infty  \hskip -1.2mm
e^{ i{N} \alpha^\prime  -iM \beta^\prime}  \vert k_1M, k_2N
\rangle \otimes \vert
-k_1N,  k_2M  \rangle \Bigr)
\end{equation}
where
\be
S^{(-)} =   S^{(+)}   \left(\hskip -1mm \begin{array}{cc}  -1 & 0 \\
                              0 & 1  \end{array}\hskip -1mm \right) \hskip -0.6mm
                                 =
                                \left(\hskip -1mm \begin{array}{cc}  {\rm cos\, 2\theta} &  -{\rm sin\, 2
                                \theta} \\
                               {\rm sin\, 2 \theta} & {\rm cos\, 2\theta}
                                 \end{array} \hskip -1mm\right)  \ , \ \ \ 
                                 \theta = {\rm tan}^{-1}\left({2k_2R_1R_2\over k_1}\right)   ,  
\ee
and the g-factor of the brane  is
\be\label{gminus}
 {g}^{(-)} = \sqrt{k_1^2 + 4 k_2^2 R_1^2R_2^2\over 4R_1R_2}
 \,  = \,   \sqrt{ k_1 k_2  \over  {\rm sin} 2\theta} 
\ .
\ee
Notice that since T-duality inverts $\sqrt{2} R_1$, the angle $\theta$ 
is in general not the same as $\vartheta$. The two angles coincide only at the self-dual point
of the radius $R_1$. 
The  state (\ref{d0d2}) describes the bound state of $k_1$ D0s 
and  $k_2$ D2s. 
As a check, note that for $(k_1,k_2) = (1,0)$ or $(0,1)$ one recovers the expressions
of  the pure D0, respectively the pure D2 brane. Note also that,  consistently with our notation, 
$-{\rm det}S^{(-)} = -1$ and the dimension of the branes is even.

\vskip 1mm
The generalization to oblique and to three-dimensional tori is straightforward.  
The boundary states for  an oblique two-torus  can be obtained by a sequence of T-duality
transformations and rotations, starting with the elementary D0 brane. Furthermore, 
starting with the general D2/D0 brane on the $(\phi^1,\phi^2)$ plane of  a three-torus, 
one  can rotate it to any other orientation in the compactification
lattice.  A  T-duality then maps this to an arbitrary D3/D1 bound state. For $c=4$ there
exist new branes (e.g. along  the Higgs branch of the D4/D0 system)  which cannot be constructed
by the above algorithm.  We will not pursue the study of these higher-dimensional
 branes
in the present work.


\section{Unfolding and the topological maps}

\subsection{The unfolding procedure}

A conformal interface between two theories,  1 and 2, can  be mapped to a
conformal boundary of the tensor-product theory ${\rm CFT}_1\otimes {\rm CFT}_2$
by the folding trick  shown in figure 1.  Conversely, we can unfold a boundary state
to an interface  whenever    the bulk CFT has  two non-interacting    
components.
Let us assume that in some  appropriate basis,  constructed by acting with 
   \vskip 8mm

\begin{figure}[h!]
\begin{center}
\includegraphics[height=3.6cm]{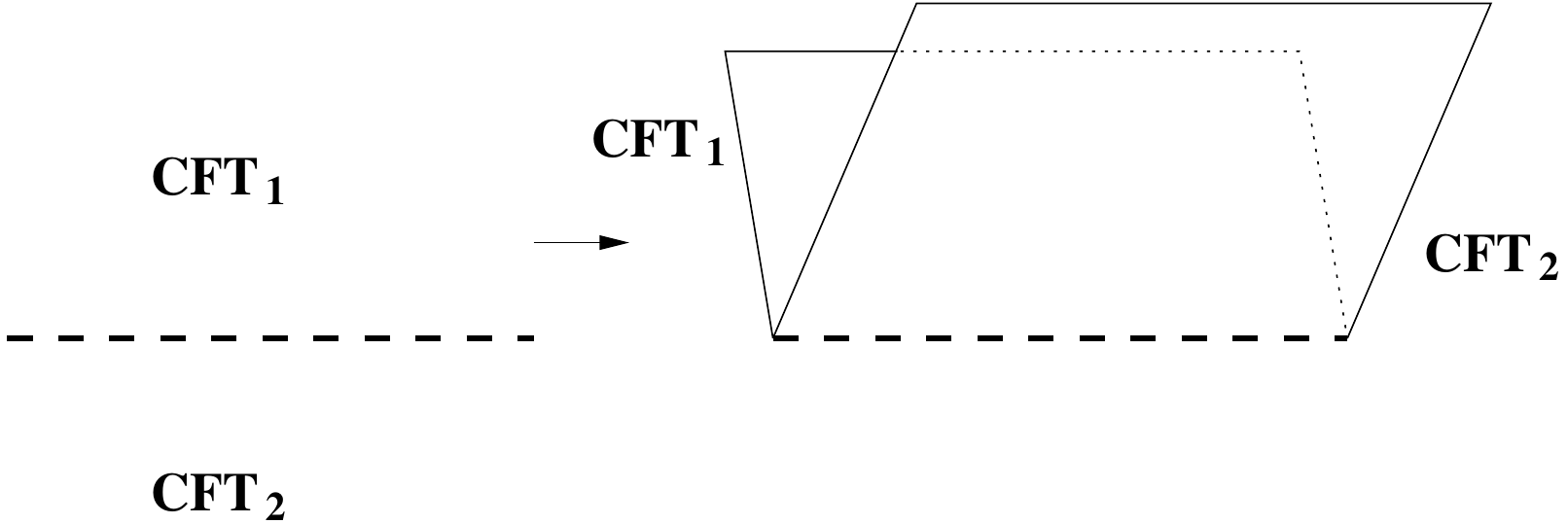}
\caption{\small Folding trick.}\label{foldgraph}
\end{center}
\end{figure}

 \noindent     self-adjoint
(left and right) operators ${\cal O}_{\lambda_j}$ and ${\cal O}_{\tilde\lambda_j}$
on the vacuum,  the boundary state takes the general form
\begin{equation}\label{general}
\vert\hskip -0.5mm \vert \, {\cal B}\, 
\rangle\hskip -0.6mm\rangle  \ =\   \sum  {\cal B}_{\lambda_1 \tilde\lambda_1  \lambda_2 \tilde\lambda_2}\, 
\vert \lambda_1,\tilde\lambda_1\rangle\otimes \vert \lambda_2,\tilde\lambda_2\rangle
\end{equation}
with coefficients ${\cal B}_{\lambda_1 \tilde\lambda_1  \lambda_2 \tilde\lambda_2}$
which are {real}.\footnote{Put differently,  the one-point functions of hermitean
bulk operators on the disk must be real, an assumption that is certainly true
for the toroidal branes
considered here.}
We also assume that both CFTs are left-right symmetric.
Then unfolding the boundary state (\ref{general}) 
leads to the following interface and anti-interface, expressed as operators
from the Hilbert space  of ${\rm CFT}_2$  to the Hilbert space of ${\rm CFT}_1$ 
and  vice-versa,  
\begin{eqnarray}
&{\cal I}{(1\leftarrow 2)} = \sum\,   {\cal B}_{\lambda_1 \tilde\lambda_1  \lambda_2 \tilde\lambda_2}\, 
\vert \lambda_1,\tilde\lambda_1\rangle\langle \tilde\lambda_2, \lambda_2\vert \ , 
\cr\   \cr 
&{\cal I}{(2\leftarrow 1)} = \sum\,   {\cal B}_{\lambda_1 \tilde\lambda_1  \lambda_2 \tilde\lambda_2}\, 
\vert \lambda_2,\tilde\lambda_2\rangle\langle \tilde\lambda_1, \lambda_1\vert\ .  
\end{eqnarray}
Notice that unfolding flips the sign of the (closed-string) time coordinate $\tau$ for the unfolded theory, say ${\rm CFT}_2$.
  It therefore  involves
both hermitean conjugation and the exchange of left- with
 right-movers,  $\lambda_2\leftrightarrow \tilde\lambda_2$. 
The individual matrix
elements of  the above operators  are two-point 
functions on the sphere in the presence of the conformal  interface.
\vskip 1mm

Let us now specialize to the D1 and D2/D0  branes
of the previous section.  Since the torus is orthogonal and there is no $B$-field background,
the two scalar fields are decoupled in the bulk, and the boundary states  can be unfolded.  
Flipping the sign of $\tau$ in the expression (\ref{phi}) sends
\be
\hat N \to -\hat N , \ \ \ a_n\to -\tilde a_n^\dagger , \ \ \ {\rm and}
\ \ \  \tilde a_n \to -a_n^\dagger\ . 
\ee
This is, as argued, hermitean conjugation followed by the exchange of left and right movers
(the minus sign  can be absorbed in the definition of basis). 
The only subtle point concerns the choice of a real basis of states. For the  
ground states, for example, one  
must work with the  basis of  states
\be
{\vert p, w\rangle + \vert -p, -w \rangle\over \sqrt{2}}\ \ \ {\rm   and}
\ \ \ {\vert p, w\rangle -\vert -p, -w\rangle\over \sqrt{2}i}\ , 
\ee
in which the coefficients $ {\cal B}_{\lambda_1 \tilde\lambda_1  \lambda_2 \tilde\lambda_2}$ are real.   
 Hermitean conjugation  followed by the reflection of momentum
 for the scalar $\phi^2$,  can then be shown to map 
$\vert p_2, w_2\rangle$ to $\langle -p_2, w_2\vert$ in the expressions  
(\ref{k1k2}) and (\ref{d0d2}) of the boundary states.  
The final result for the interface operators therefore
reads:
 \be\label{Imain}
{\cal I}^{(\pm)\ (R_1\leftarrow R_2)}_{(k_1,k_2)} =\,   {L}^{(\pm)}_{(k_1,k_2)} 
\prod_{n=1}^\infty e^{\left( S^{(\pm)}_{11} (a_n^1)^\dagger
(\widetilde a_n^1)^\dagger - S^{(\pm)}_{12} (a_n^1)^\dagger a_n^2 - S^{(\pm)}_{21}
 (\widetilde a_n^1)^\dagger \widetilde a_n^2 +S^{(\pm)}_{22} a_n^2 \widetilde a_n^2
 \right) }\  ,  
\ee
where the ground state operators are the lattice sums:
 \begin{equation}\label{k1k2i}
{L}^{(+)}_{(k_1,k_2)}( \alpha,  \beta)  
\,   =   \  \,  \sqrt{ k_1 k_2 \over {\rm sin} 2\vartheta } 
  \times\hskip -0.8mm  \sum_{N,M =-\infty}^\infty  
e^{i{N} \alpha -iM \beta}  \vert k_2N, k_1M\, 
\rangle \langle 
k_1N,  k_2M  \vert  \ , 
\end{equation}
 \begin{equation}\label{d0d2i}
 {L}^{(-)}_{(k_1,k_2)}( \alpha,  \beta)  \,  = \,  \,  \sqrt{k_1 k_2 \over {\rm sin} 2\theta } 
  \times     \sum_{N,M =-\infty}^\infty  
e^{i{N} \alpha  -iM \beta}  \vert k_1M, k_2N\, 
\rangle \langle 
k_1N,  k_2M  \vert  \ ,  
\end{equation}
 and  in eq. (\ref{Imain}) the daggered oscillators act  implicitly on $L^{(\pm)}$ from the left. 
 For the reader's convenience, we collect here also the expressions for $S^{(\pm)}$:
  \begin{equation}\label{S}
 S^{(+)}  =   \left(\hskip -1mm \begin{array}{cc}  -{\rm cos\, 2\vartheta} &  -{\rm sin\, 2\vartheta} \\
              - {\rm sin\, 2\vartheta} & {\rm cos\, 2\vartheta}  \end{array} \hskip -1mm\right)  \ ,   
              \ \ \   \vartheta = {\rm tan}^{-1} ({k_2R_2\over k_1R_1})\ ,                          
 \end{equation}
\be\label{finS}
S^{(-)} =    \left(\hskip -1mm \begin{array}{cc}  {\rm cos\, 2\theta} &  -{\rm sin\, 2
                      \theta} \\
                    {\rm sin\, 2 \theta} & {\rm cos\, 2\theta}
                      \end{array} \hskip -1mm\right)  \ , \ \ \ 
                \theta = {\rm tan}^{-1}\left({2k_2R_1R_2\over k_1}\right)  .  
\ee
Eqs. (\ref{Imain})  to (\ref{finS}) define the most general interfaces which separate two free-boson theories with radii  $R_1$ and  $R_2$, and which  preserve a $U(1)\times U(1)$ subgroup of the $U(1)^4$ symmetry of the bulk.  Below we will refer to the $+$ and the $-$ operators as,
respectively, even and odd. When no confusion is possible, their
 dependence on the phases $(\alpha , \beta)$ and on the radii $(R_1\leftarrow  R_2)$ will be omitted.  
\vskip 2mm


\subsection{Reflection, transmission and entropy}

It is important here to note that the operators  (\ref{Imain}) depend on the radii  only through
the angles $\vartheta$  and $\theta$.  This is true in particular for the matrices $S^{(\pm)}$,  
whose  elements  are the reflection and transmission coefficients across the  
interface \cite{bbdo,Quella2}.  Total reflection occurs when $\vartheta$ or
$\theta$ is a multiple of $90^o$, which
requires either $k_1$ or $k_2$ to vanish. This corresponds 
(up to Chan-Patton multiplicity) to the four factorizable boundary states of section 2,  
which unfold to the  interface operators
\be
{\cal I}_{\rm refl}\  =\  \vert\hskip -0.5mm \vert \, {\rm D}r_1 \rangle \hskip-0.6mm \rangle
 \langle \hskip-0.6mm \langle  D{r_2}  \vert\hskip -0.5mm \vert \ \ \ \ 
 {\rm with}\ \ \ \  r_1,r_2=0,1\ . 
\ee
The two CFTs have  in this case separate consistent boundaries, and they decouple
completely.

More interesting is the case of total transmission, which occurs for
 angles that are an odd  multiple of  $45^o$.  The  interface operators   have now  the form
\be\label{formtopo}
{\cal I}^{(\pm)}_{\rm top} = \  L^{(\pm)}\,  \prod_{n=1}^\infty  {e}^{(-)^l \left[
 (a_n^1)^\dagger a_n^2 \pm  (\widetilde a_n^1)^\dagger \widetilde a_n^2\right]} 
\, , \ \ \ \  {\rm for}\ \ \ \  \vartheta ,\theta = (2l+1){\pi\over 4}\, . 
\ee
It follows  that the energy-momentum tensor is continuous across
the interface, i.e. the Virasoro generators (not to be confused with the lattice sums!) 
obey the commutation relations
\be\label{comm}
{\rm L}^1_n\,  {\cal I}^{(\pm)}_{\rm top} \, =\,  {\cal
  I}^{(\pm)}_{\rm top}\,
  {\rm L}^2_{n}\ \ \ {\rm and}
\ \ \ \  \widetilde {\rm L}^1_n\,  {\cal I}^{(\pm)}_{\rm top} \, =\, 
 {\cal I}^{(\pm)}_{\rm top}\,  \widetilde {\rm L}^2_{n}\ . 
\ee 
Such interfaces have been dubbed topological  
, because  they can be deformed freely  across a 
Riemann surface.  The topological   interfaces for
toroidal  CFTs  (both symmetric and non-symmetric) 
were analyzed recently in ref. \cite{mat2}.  Here we will  only discuss  a few,  
relevant  for our purposes,  features. 
\vskip 1mm

Consider first the case of defects, i.e. $R_1=R_2=R$.  As argued generally  
by Fr\"ohlich et al  \cite{juerg},    
the topological defects should include the generators of automorphisms of the CFT. 
At a generic value of the radius  the only topological defects are
    \be\label{top1}
 e  (\alpha,\beta) \equiv  {\cal I}^{(+)\ (R\leftarrow R)}_{(1,1)} \  \ \ \   {\rm and}
  \ \ \ \    {r}(\alpha, \beta) \equiv  {\cal I}^{(+)\ (R\leftarrow R)}_{(1,-1)}\ .  
\ee
These generate indeed  the  semidirect
product $U(1)^2\rtimes Z_2$,  i.e. the left 
and right translations and the reflections of  $\phi$.
Notice that the trivial (identity) defect 
is $e(0,0)$,  i.e.   a diagonal D1-brane
in the $(\phi^1, \phi^2)$ plane, after folding. Turning on a Wilson line, translating
and/or reflecting this diagonal D1-brane,  gives all the other symmetry generators
for generic $R$. 
At the self-dual radius,  $2R_*^2=1$,   there appear two new topological defects, 
 \be\label{top2}
 \omega \equiv  {\cal I}^{(-)\ R_*\leftarrow R_*}_{(1,1)}  
 \  \ \ \   {\rm and}
  \ \ \ \    {\tilde \omega}  
 \equiv 
 {\cal I}^{(-)\ R_*\leftarrow R_*}_{(1,-1)} \ . 
\ee
These generate T-duality twists,  i.e. separate left and right reflections of $\phi$.  
They enhance the symmetry to  $(U(1)\rtimes Z_2)^2$, which is the subgroup 
that preserves a maximal torus of the full $SU(2)^2$ symmetry of the self-dual  theory. 
The missing generators break more than two of the original $U(1)$ symmetries
of the free-scalar model, which
explains why  they were not
included in our set of defects. Thus our analysis agrees with  the observation of 
ref.   \cite{juerg}.

\vskip 1mm    

What about other topological interfaces and defects?  From the expression for the angles
one sees that  (provided $k_1k_2\not= 0$) 
every one of the operators (\ref{Imain}) 
becomes  topological at a special
value of the ratio,  or of the product of radii. Specifically this happens  when
\be\label{top22}
{R_2\over R_1} =   \Bigl\vert {k_1\over k_2}\Bigr\vert \ \ \ \ {\rm or}\ \ \ \ \ \ 
2 R_1R_2 = \Bigl\vert {k_1\over k_2}\Bigr\vert
\ee
in the even, respectively odd case. Inspection 
of the lattice sums  (\ref{k1k2i}) and (\ref{d0d2i}) 
reveals that when $\vert k_1 k_2\vert > 1$ these operators map all but the
proper  sublattice    $\vert k_1 \mathbb{Z} , k_2 \mathbb{Z} \rangle$  of the
ground states to zero. The pairs of states that survive
in these sums are states with equal $U(1)$ charges and 
 conformal weights.  These higher topological interfaces
do not therefore correspond to invertible operators, but rather to  projectors,  coupled  with
isomorphisms  of appropriate subsectors of the two CFTs.
For example, the $(2,1)$ topological interface maps the even-winding-number
states  of a theory at radius $R$ to
 the even-momentum-number  states of a theory at radius $2R$. 

\vskip 0.5 mm

   An important feature of topological interfaces is that they minimize the entropy, 
   defined as the logarithm of the $g$ factor,  when
    the radii vary with  $(k_1, k_2)$ held fixed. This is  a property reminiscent 
    of the minimum-energy condition for  BPS states.  It is a simple consequence of the
    general expression for the $g$-function ($\vartheta$ must be 
    replaced by $\theta$ in the odd case):  
\be
{\rm log}\, g \ = \    {\rm log}\sqrt{ \vert k_1 k_2\vert}
  - \, {\rm log}\sqrt{ \vert  {\rm sin} 2\vartheta\vert}\ . 
\ee
The second contribution (which depends on the reflectivity  \cite{bbdo,Quella2} of the interface)
 is non-negative, and it vanishes only in the topological case.  
The first, irreducible contribution is also non-negative, and it vanishes only for 
 the  symmetry-generating defects $(k_1, k_2) = (1, \pm 1)$.  These  latter are  
 the only invertible maps,  which is consistent with the fact that they 
 should not generate any  entropy.  
 We conjecture that topological\footnote{More general
 interfaces can have a $g$ factor  smaller than one, and hence a negative entropy. 
An example is the totally-reflecting interface corresponding to a simple D2-brane,
for which $g= \sqrt{R_1R_2}$.   We thank Ingo Runkel and the JHEP referee for
 pointing this out.}
interfaces between unitary conformal theories always have non-negative entropy,
and that the bound is saturated only by CFT isomorphisms.
  
\vskip 0.5 mm 
 The analogy of topological interfaces 
 with BPS black holes can actually  be  pushed even further: one can  interpret
 the topological conditions (\ref{top22}) as an attractor mechanism \cite{attractor} 
 that fixes the moduli of the bulk theory for any given set of  ``charges''   
 $(k_1, k_2, \pm)$. Notice that there are two asymptotic regions and hence two
 bulk radii, but  only one combination of the two is being fixed.  Also,   the entropy
 of a topological interface
 is the sum of  the logarithms, rather than of the absolute values,  
 of the  integer charges.  This is compatible  with the fact that charges are
 multiplicatively conserved, as we are now going to explain.


\section{The algebra of  interfaces}   

\subsection{Topological reduction of the fusion}

Two boundary states,  and hence also the corresponding interface operators,
can be added. If they are identical, this amounts  simply to introducing a Chan-Patton multiplicity.
On the other hand, two operators can  also be multiplied whenever  the image of one lies in
the domain of definition of the other. In the case at hand, this corresponds to juxtaposing 
an interface
between ${\rm CFT}_1$ and ${\rm CFT}_2$ and  an interface between 
${\rm CFT}_2$ and ${\rm CFT}_3$, as shown in the figure 2. Because  the product  of these
 two
operators is in
\vskip 2mm
\begin{figure}[h]
\begin{center}
\includegraphics[height=4.5cm]{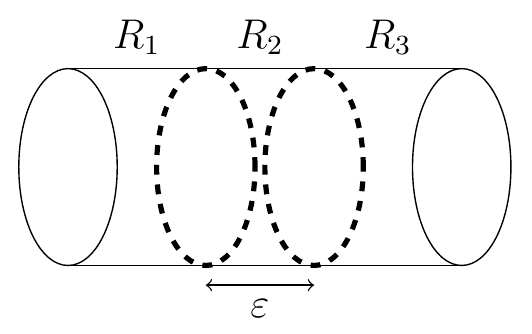}
\label{fusiongraph}
\caption{\small Fusion of two interfaces between three CFTs with radii $R_1$, $R_2$ and $R_3$. 
In the limit of vanishing separation,   $\varepsilon\to  0$, 
the result should not depend on the value of the radius in the middle region. }
\end{center}
\end{figure}

\noindent   general singular, we must first separate the interfaces by a distance 
$\varepsilon$. We work as before on
the cylinder $(\sigma, \tau)$,  so that the periodicity of $\sigma$ sets the scale of distance. 
By the usual arguments of QFT we expect  that the limit $\varepsilon\to 0$ can be rendered
finite by a local self-energy counterterm. 
 Accordingly, we define the fusion of the two interfaces as follows:
 \be\label{compp}
  {\cal I}\circ  {\cal I}^\prime \ 
  \equiv \   {\rm lim}_{\varepsilon\to 0} \  \ e^{{2\pi} d/\varepsilon}\
 {\cal I}{(1\leftarrow 2)} \,  e^{-\varepsilon H_2}\,  {\cal I}^\prime{(2\leftarrow 3)} \ , 
\ee
where $H_2$ is the generator of $\tau$-translations in  ${\rm CFT}_2$,  and 
$d/\varepsilon$ is the self-energy counterterm which  must be
adjusted so as to make the right-hand-side finite. Notice that this procedure 
is unambiguous because $\varepsilon$
is the only relevant length scale in the problem.\footnote{The  inverse ``temperature'' $2\pi$ 
may  only appear  multiplicatively in the exponent.}
 The fact that  ${\cal I}\circ  {\cal I}^\prime$ 
 should be  the sum of  elementary $(1\leftarrow 3)$ interfaces with  integer
 non-negative coefficients  is a non-trivial check  of the consistency of this definition.
   
  Now the following 
intuitive  but naive
argument motivates  one of the main points of this paper: 
in the limit $\varepsilon\to 0$  the region in the middle disappears, and so should any
memory of the value of the radius in this region. Thus, modulo a local renormalization,
the result should be independent of the value of $R_2$.  
To be more precise, given interfaces ${\cal I}_{(k_1,k_2)}^{(R_1 \leftarrow R_2)}$ and ${{\cal I}'}_{(k_1',k_2')}^{(R_2 \leftarrow R_3)}$
the fusion product is expected to be independent of variations in $R_2$, while
all other quantities $R_1,R_3, k_i, k_i'$ are held fixed.
An additional argument is provided  by continuity:  if the fusion coefficients
are integers   they should not jump around as we vary $R_2$, except possibly at
singular points in the moduli space. We will actually show  that these naive arguments  are correct 
in  the case at hand, by computing explicitly  (\ref{compp}) 
in the following section. 
Here we assume the result,  and proceed to 
 calculate the algebra. 

\vskip 1mm  
 This is made easy by the following trick: since the 
  value of $R_2$ is irrelevant, we may choose  it  so as  to make  the interface ${\cal I}^\prime$  
topological.  We have seen in the previous section that this is always possible, as long as 
   $k_1^\prime k_2^\prime$ does
  not vanish.\footnote{If   $k_1^\prime k_2^\prime =0$, 
   then ${\cal I}^\prime$ is totally reflecting  and the CFT3
decouples. The problem reduces to the fusion of an interface with a boundary,
a case that we will discuss  in the end.} 
  Now using the commutation property of topological interfaces, eq.  ({\ref{comm}),
 we  find 
\be\label{36}
{\cal I}\,  e^{-\varepsilon H_2} {\cal I}^\prime_{\rm top} =  
 {\cal I}\,  {\cal I}^\prime_{\rm top} \, e^{-\varepsilon H_3}  ,
\ee
and on the right-hand-side  the  $\varepsilon\to 0$ limit  
can be taken smoothly. 
Put differently, once ${\cal I}^\prime$ has been
made topological, it can be moved at no cost.
 We may thus restrict attention to non-singular 
 products  of one topological with  one arbitrary interface.

   The following observation simplifies
 the calculation   
  even further:  let us define the basic radius-changing interface, which
 is the deformed identity operator 
 \be
 e_{\rm def}^{(R_1\leftarrow R_2)}  \, \equiv \,  {\cal I}^{(+)\ (R_1\leftarrow R_2)}_{(1, 1)} \ \ 
 \ \ {\rm with}\ \ \ \  \alpha=\beta = 0\ . 
 \ee
Now an arbitrary conformal interface can be obtained as the product of a topological
interface with this basic deformed identity. Explicitly: 
\be\label{eqdeforme}
{\cal I}^{(\pm)\ (R_1\leftarrow R_2) }_{(k_1, k_2)}\   =  \ 
  {\cal I}^{(\pm)\ (R_1\leftarrow R)}_{(k_1, k_2)} \  e_{\rm def}^{(R\leftarrow R_2)}\  =
   e_{\rm def}^{(R_1\leftarrow R^\prime)}  {\cal I}^{(\pm)\ (R^\prime\leftarrow R_2)}_{(k_1, k_2)} \ \ , 
  \ee
where $R$ and $R^\prime$
are here implicitly  adjusted so as to make the $(k_1, k_2)$ operators topological. 
Let us prove the first  equality, by evaluating explicitly
the product in the even case and with  $k_1k_2>0$
(the other three cases work in a similar way).  From the general form
 (\ref{formtopo}) we   see that
  the topological $(R_1\leftarrow R)$ operator  commutes with the oscillator  modes,
 \be
 a_n^1\,  {\cal I}^{(+)}_{\rm top} = {\cal I}^{(+)}_{\rm top} \, a_n \ \ \ \ \ {\rm and}\ \ \ \ \
 \tilde a_n^1\,  {\cal I}^{(+)}_{\rm top} = {\cal I}^{(+)}_{\rm top}\,  \tilde a_n\ ,  
\ee
where $a_n$ and $\tilde a_n$ refer to the region of radius $R$,
 and the same equations hold for daggers. Thus in the expression
  (\ref{Imain}) for the basic $R\leftarrow R_2$ interface we may replace 
the $a_n^\dagger$ by $(a_n^1)^\dagger$,  and  the $a_n^\dagger$ by $(a_n^1)^\dagger$.
Furthermore the  angle that enters 
the $S$-matrix of this interface  is given by
 ${\rm tan}\vartheta = R_2/R= k_2R_2/k_1R_1$, 
where the second step uses the topological property of the $R_1\leftarrow R$ operator. 
Put differently, detaching a topological part does not change the reflectivity of the
interface. 
Finally,   the lattice sum in $e_{\rm def}$ is just the trivial isomorphism of 
momentum and winding  states. Multiplying
 with the lattice sum of the $R_1\leftarrow R$ operator completes the construction of
 the $R_1\leftarrow R_2$ operator,  and proves
the relations  (\ref{eqdeforme}) . 
 
\vskip 1mm

These relations  show that all  conformal interfaces (\ref{Imain}) can  
be written as products of a deformed identity and a topological ``dress'', 
and that the latter can be moved off to the left or right. 
We can therefore calculate the product  (\ref{36})  by stripping 
  ${\cal I}$ of its  topological  dress, 
  multiplying  this with the operator  ${\cal I}_{\rm top}^\prime$,  and  then  dressing back  
  the deformed identity  on the left side. 
 Hence,  we need only  to study the  products of topological operators.

\subsection{The multiplicative law for the charges}

 Since the oscillator
modes enter in such products trivially,   it is sufficient  to multiply their  lattice sums. 
Furthermore, by acting with the symmetry generator $e(\alpha,\beta)$ from the left
or the right, we may set all  phases in these lattice sums to zero. Notice that this
 is a non-commutative operation, e.g.  
\be\label{eg}
 {\cal I}_{(k_1, k_2)}^{(+)}(\alpha, \beta)\  = \ 
e({\alpha\over k_2} ,  {\beta\over k_1} )\  {\cal I}_{(k_1, k_2)}^{(+)}(0, 0)\  = \ 
 {\cal I}_{(k_1, k_2)}^{(+)}(0, 0)\  e({\alpha\over k_1} ,  {\beta\over k_2} )\,  
\ee
with a similar equation for the odd case. When  all the phases are set to zero,  
the product of  two even topological operators reads: 
 \be\label{lower}
\sqrt{\vert
k_1k_2k_1^\prime k_2^\prime\vert}\,  \sum_{N,M, N^\prime, M^\prime} \vert k_2N, k_1M
\rangle \langle 
k_1N,  k_2M \,  \vert\vert\,  k_2^\prime N^\prime, k_1^\prime M^\prime
\rangle \langle 
k_1^\prime N^\prime,  k_2^\prime M^\prime  \vert 
\nonumber 
\ee
 \be 
= \  J J^\prime \sqrt{\vert
K_1K_2 \vert } \,  \sum_{N,M} \vert J K_2 N,   J^\prime K_1M
\rangle \langle \, 
 JK_1  N , \, J^\prime K_2  M \,  \vert \ , 
\ee
where in the lower line we have redefined $N$ and $M$ so that they run unconstrained
over all the integers, and
 \be
 J = {\rm gcd}(k_1^\prime, k_2)  \ , \ \  J^\prime = {\rm gcd}(k_1, k_2^\prime)\ ,\ \ 
 K_1 = {k_1 k_1^\prime\over JJ^\prime}\ , \ \ K_2= {k_2 k_2^\prime\over JJ^\prime} \ , 
 \ee
with  ``gcd'' standing  for the greatest common divisor.  
 If  $J=J^\prime =1$,  the product  is  just  the elementary
$(K_1, K_2)$ interface.    More generally, it is an array of $JJ^\prime$ such interfaces,
arranged periodically in the $(\alpha, \beta)$ parameter space \cite{mat2}. 
Periodic arrays  couple   indeed only to a sublattice of momenta and windings, as the reader
will have no difficulty to verify.  
Explicitly, the  product formula  reads
\be\label{algebra}
{\cal I}^{(+)}_{(k_1, k_2)}(0,0)  \circ  {\cal I}^{(+)}_{(k_1^\prime, k_2^\prime)}(0,0)
=  \sum_{j, j^\prime} \,    {\cal I}^{(+)}_{(K_1, K_2)} ({2\pi j\over J} ,
 {2\pi j^\prime\over J^\prime})\ , 
\ee
where the sums run over $j = 1, \cdots, J$ and $j^\prime = 1, \cdots, J^\prime$. 
\vskip 1mm  


This result can be expressed more elegantly  if we mod out the action 
of the $U(1)^2$ symmetry. Let us denote by  $[ k_1, k_2]^{(+)}$
the equivalence class of all D1-branes winding $(k_1, -k_2)$ times around the
$(\phi^1, \phi^2)$ torus. We also  relax the condition that the winding numbers be
relatively prime, but take note of the fact that the (open-string) moduli space
has dimension equal to  $2\, {\rm gcd}(k_1, k_2)$. 
For these equivalence classes of D-branes  the fusion formula takes the simple form
\be\label{equivc}
[ k_1, k_2]^{(+)}\circ [ k_1^\prime, k_2^\prime]^{(+)}\ 
=\ [ k_1k_1^\prime, k_2k_2^\prime]^{(+)}\ . 
\ee
As anticipated already  in section 3,  the interface charges get multiplied and  
 topological fusion  conserves the entropy. 
Notice that  the dimension  of the moduli space of the product can be 
bigger than the sum of dimensions of its two factors. Thus  a generic representative
in the equivalence class on the right-hand side 
need not factorize. For example, three elementary 
$(1,1)$ interfaces can only be written in the
product form $(1, 3)\circ (3, 1)$ if they are arranged in a periodic array. 
  
To complete the derivation of the algebra, we need also  to analyze  the odd case.
This can be done with the help of the T-duality twist,  defined in  eq. (\ref{top2}) 
at the self-dual point.  Clearly this operator remains topological for any pair of
radii such that  $2R_1R_2=1$.   A simple
calculation shows that $\omega$  squares to 1, and that 
it exchanges the even and the odd interfaces  as follows:
\be\label{twodd}
\omega \circ {\cal I}^{(-)}_{(k_1, k_2)}(\alpha, \beta)  =
 {\cal I}^{(+)}_{(k_1, k_2)} (\alpha, \beta)\, 
 \ , \ \ \ 
  {\cal I}^{(-)}_{(k_1, k_2)}(\alpha, \beta) \circ \omega
   = {\cal I}^{(+)}_{(k_2, k_1)} (-\beta, -\alpha)\, . 
\ee
 Together with equations (\ref{eg}) and  (\ref{algebra}),  these twist relations 
are  sufficient to compute the fusion of 
any two  interfaces in the list (\ref{Imain}). 
The final result,   generalizing (\ref{equivc}),   can be
worked out easily: 
 \be\label{equivc1}
[ k_1, k_2]^{(\pm)}\circ [ k_1^\prime, k_2^\prime]^{(+)}\ 
=\ [ k_1k_1^\prime, k_2k_2^\prime]^{(\pm)}\ , \nonumber 
\ee
\be\label{equivc-}
[ k_1, k_2]^{(\pm)}\circ [ k_1^\prime, k_2^\prime]^{(-)}\ 
=\ [ k_2 k_1^\prime , k_1k_2^\prime]^{(\mp)}\ . 
\ee
A simple corollary of the above fusion rules is that the symmetry generators, together
with one representative in the $[1, p]^{(+)}$  class for each prime number $p$, are sufficient 
to generate the entire   algebra of  these  topological  interfaces. 

\vskip 1mm
 These fusion relations  continue to hold for
 totally-reflecting interfaces, i.e. in the special case  
 $k_1^\prime k_2^\prime =0$.  They then describe the action of the
interface operators on the D-branes of the  $c=1$  model. For example, 
a Dirichlet condition in the left-half space corresponds to an operator in the class $[1, 0]^{(-)}$
or $[0, 1]^{(+)}$, 
where the  two choices differ by a twist in the decoupled right-half region.
Acting on  this D0-brane with
 an operator in the $[k_1, k_2]^{(+)}$ class  produces a periodic
 array of D0-branes, as illustrated in figure 3. This agrees

  \begin{figure}[h!]
\begin{center}
\hskip 1cm \includegraphics[height=4.5cm, angle=270, scale =2.2]{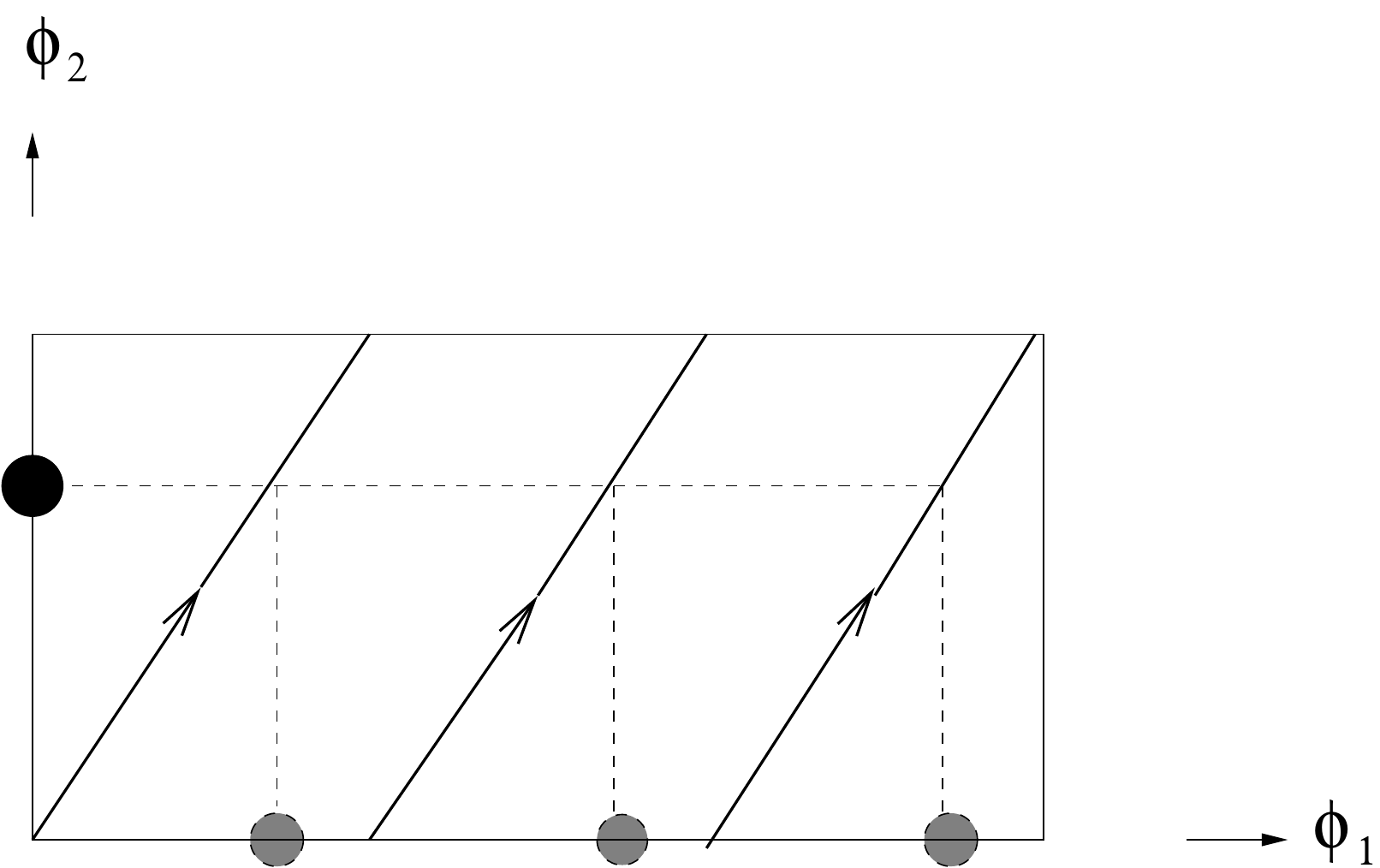}
\vskip 3mm
\caption{\small  The action of a $[1, 3]^{(+)}$ operator on the 
D0-brane of  theory 2 (black dot)  produces three D0-branes of theory
1 (light-colored  dots). The latter are arranged
periodically around the circle. }\label{mapgraph}
\end{center}
\end{figure}

\noindent   with the simple geometric
 rule that was sketched  in the introduction. All other actions of our 
 interface operators on
 the D0-brane  and the D1-brane  of the $c=1$ model can be obtained from this
 picture by T-duality twists.  
 This completes our discussion of the algebra. 
 We turn now to a proof of the argument
that allowed the reduction to the topological case.


\section{Entropy release and stability}  

\subsection{Proof of the topological reduction}

Let us return to the definition (\ref{compp}) of the fusion product. 
Using the   relations  (\ref{eqdeforme}) we can strip  off  the non-trivial 
topological parts, if any, 
of ${\cal I}$ and ${\cal I}^\prime$ to the left, respectively right, leaving in
the middle two deformed identities, i.e. two basic radius-changing interfaces
in the $(1,1)$ sector. This is illustrated in figure 4. The
 two radii,  $R_1^\prime$ and $R_3^\prime$,  in the nucleated regions are
fixed  by the requirement that the split-off interfaces be topological, as was
explained

\vskip 3mm

 \begin{figure}[h!]
\begin{center}
\includegraphics[height=2.8cm]{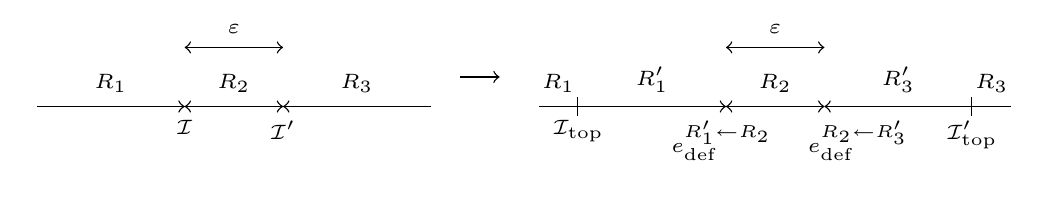}
\caption{\small By stripping off their topological parts,  we can relate the 
singularity in  the fusion of any two conformal interfaces to the singularity in
the product  of two basic radius-changing  operators. }
\end{center}
\end{figure}

\vskip 2mm

\noindent   in  the previous section. We may now take
 the limit $\epsilon\to 0$, before  dressing  back the result  with the topological interfaces
 from  the left and right.  To prove our claim,  it is  therefore sufficient to show   that
 for any  arbitrary triplet of radii we have
 \be
 e_{\rm def}^{(R_1^\prime\leftarrow R_2)}\circ e_{\rm def}^{(R_2\leftarrow R_3^\prime)}
 = e_{\rm def}^{(R_1^\prime\leftarrow R_3^\prime)}\  , 
 \ee 
i.e. that the product of deformed identities is the deformed identity. 
In the rest of this section we will explain why this is true.  Readers not interested
in the technical details can jump ahead to the next subsection. 

\vskip 0.5mm
  
 Because the different frequencies of the scalar field do not talk, the calculation 
 of the product of two interfaces
   factorizes into a separate calculation in each 
 frequency sector. Thus the product,  before taking the coincidence  limit,   reads
  \be\label{beforelimit}
  e_{\rm def}^{(R_1^\prime\leftarrow R_2)}\ q^{H_2}\  e_{\rm def}^{(R_2\leftarrow R_3^\prime)}
  =
  {1\over \sqrt{  {\rm sin}\, 2\vartheta^\prime \ {\rm sin}\, 2\vartheta } }\, 
  \sum_{N,M} q^{h_{N,M}}  \vert N, M\rangle \langle N, M\vert \, \prod_{n=1}^\infty  O_n  \ , 
 \ee
where $q \equiv e^{-\epsilon}$,  $h_{N,M}$ is the energy of the $(N,M)$ ground state in
  ${\rm CFT}_2$,  the operators $O_n$ are the result of evaluating the product  
   in the $n$th-frequency sector, and we have defined the  
   angles 
\be
{\rm tan}\vartheta = {R_2\over R_1^\prime}\ , \ \ 
{\rm tan}\vartheta^\prime  = {R_3^\prime \over R_2}\ \ \ \ {\rm and\ }\ \ \ \ 
{\rm tan}\Theta = {R_3^\prime \over R_1^\prime}\ = {\rm tan}
\vartheta^\prime\, {\rm tan}\vartheta\ . 
\ee 
Notice that,  as we have stressed earlier,  the topological dressing of an interface does not 
affect  its  angular orientation.  
The operator  $O_n$ is the product of  the $n$th-frequency exponentials  in the
general expression (\ref{Imain}) for a  conformal interface, 
with $q^{H_2}$ sandwiched in the middle, and with the whole thing 
evaluated in the ground state of ${\rm CFT}_2$. The
result depends on the oscillators
 $(a_n^1)^\dagger$, $(\tilde a_n^1)^\dagger , a_n^3$ and $\tilde a_n^3$ in the outer regions,
 as well as on the evolution parameter
 $q$ and on the angles $\vartheta$ and $\vartheta^\prime$.   
  
\vskip 1mm

To do this calculation, note that the oscillators of the outer regions can be treated
effectively as c-numbers, and that the evolution operator can be absorbed into
a rescaling of the daggered oscillators of the middle region,
\be
(a_n^2)^\dagger \to q^n (a_n^2)^\dagger\ \ \ {\rm and}\ \ \ 
(\tilde a_n^2)^\dagger \to q^n (\tilde a_n^2)^\dagger\ . 
\ee
 To lighten the notation, we will  replace below  $(\tilde a_n^2)^\dagger$ by  $a^\dagger$ 
 and similarly for the tilde oscillators. We also use the shorter notation
 $c \equiv {\rm cos\, 2\vartheta}$, $s \equiv {\rm sin}\, 2\vartheta$ and similarly for  
 $\vartheta^\prime$. From eqs. (\ref{Imain}) and   (\ref{S}) we can now read off the
 following expression for the operator $O_n$  in the $n$th sector:  
 \be\label{matr}
 O_n =  e^{B_1 +B_3}  \langle 0\vert e^{ \left( c\,  a \tilde a + a A_1 
  + \tilde a \tilde A_1\right)  } \  e^{\left( - q^{2n} c^\prime 
 a^\dagger \tilde a^\dagger + q^n a^\dagger A_3 + q^n 
 \tilde a^\dagger \tilde A_3\right) }\vert 0\rangle
 \ , 
 \ee
where  $\vert 0\rangle$ is the ground state of the $(a, \tilde a)$
system,
 and the capital letters
stand for the following mutually-commuting operators:
\be
A_1 = s\,  (a_n^1)^\dagger\ , \ \ \   \tilde A_1 = s\,  (\tilde a_n^1)^\dagger\ ,
\ \  B_1 = -c \, (a_n^1)^\dagger (\tilde a_n^1)^\dagger\nonumber 
 \ee
 \be
 A_3 = s^\prime\, a_n^3\ , \ \ \ \ \tilde A_3 = s^\prime\, \tilde a_n^3\ , \ \ \ 
 B_3 = c^\prime\, a_n^3 \tilde a_n^3\ . 
 \ee
We can calculate the matrix element in (\ref{matr}) by  
 using the Gaussian representation, eq.  (\ref{gauss}),  and 
 commuting the order of the exponentials so that  $a$ passes to the right of $a^\dagger$.
  This is similar to the annulus  calculation done in   section 2.
 The result reads
  \be
 O_n =  e^{B_1 +B_3}  \int {d^2z d^2w\over \pi^2}  \,  e^{-z\bar z - w\bar w}  
 e^{(A_1 +  c z)(q^n A_3 -  q^{2n} c^\prime w)} 
 e^{(\tilde A_1 + \bar z )( q^n \tilde A_3 + \bar w)} \ . 
 \ee
Performing  the Gaussian integrations over $z$ and $w$ and  doing some straightforward
algebra  gives :
\be
O_n \ = \  {1\over 1+ c c^\prime q^{2n} } \ {\rm exp} \left[ 
  \left(  (a_n^1)^\dagger\ \  \tilde a_n^3 \right) 
 \left(\hskip -1mm \begin{array}{cc}  M_{11} &   -M_{12}\\
             -M_{21}  & M_{22}  \end{array} \hskip -1mm\right) 
 \left( \hskip -1mm  \begin{array}{c}   (\tilde a_n^1)^\dagger \\
             a_n^3  \end{array} \hskip -1mm  \right)   \right]   \nonumber      
\ee
 with
 \be
 M = {1\over 1+ c c^\prime q^{2n} } \  \left(\hskip -1mm
  \begin{array}{cc}   -c - c^\prime q^{2n} &    s s^\prime q^n \\
         s s^\prime q^n   &  c^\prime + c q^{2n}  \end{array} \hskip -1mm\right)  \ .    
 \ee
Plugging this result in eq. (\ref{beforelimit}) gives the final expression for the
product of two basic interfaces
separated by a distance $\varepsilon =  -{\rm log}q$\ . 
 
 \vskip 1mm 
We are now ready to take the $q\to 1$ limit.  Simple trigonometry shows that
in this limit $M$ goes over smoothly to $S^{(+)}(\Theta)$, where  
$\Theta$ is the angle of the basic $(1\leftarrow 3)$ interface.
Furthermore the lattice sum goes over smoothly to a multiple of the identity operator. 
 Thus in the end
 \be
 e_{\rm def}^{(R_1^\prime\leftarrow R_2)}\circ e_{\rm def}^{(R_2\leftarrow R_3^\prime)}
 \ =   \ {\cal N}\ e_{\rm def}^{(R_1^\prime\leftarrow R_3^\prime)}\  , 
 \ee 
 where the normalization constant reads
 \be
 {\cal N} \ = \  {\rm lim}_{\varepsilon\to 0} \  \ e^{{2\pi} d/\varepsilon}\ \sqrt{{\rm sin}\, 2\Theta
 \over {\rm sin}\, 2\vartheta\  {\rm sin}\, 2\vartheta^\prime}\ 
 \prod_{n=1}^\infty (1 + {\rm cos}\, 2\vartheta
 \ {\rm cos}\, 2\vartheta^\prime  q^{2n})^{-1}  \ ,  
 \ee
and $d/\varepsilon$ is the divergent self-energy counterterm. To calculate the product
in the limit, we use the
Euler-MacLaurin formula: 
\be\label{euler}
- \sum_{n=1}^\infty {\rm log} (1 + c c^\prime e^{- 2 n \varepsilon})  = - 
{1\over 2\varepsilon} \int_0^1 {dx\over x}\,  {\rm log}( 1 + c c^\prime x)  + 
 {1\over 2} {\rm log} (1+ c c^\prime) + \cdots 
\ee
The divergent part  was first computed in ref. \cite{bbdo}. It is a  Casimir energy,
which is here removed by  $d/\varepsilon$. 
The subtraction is,  as we explained,  unambiguous because $\varepsilon$ is
the only ultraviolet scale of the problem.
  The remaining finite terms combine nicely,  with the help of the trigonometric identity
\be
 {\rm sin}\, 2\Theta\  = \  {{\rm sin}\, 2\vartheta\ {\rm sin}\, 2\vartheta^\prime\over
 1 + {\rm cos}\, 2\vartheta\  {\rm cos}\, 2\vartheta^\prime}\ , 
\ee  
  to give ${\cal N} =1$. This completes the proof  that the fusion is independent
of the radius $R_2$ in the squeezed-in region, as advertized. 
\vskip 1.5mm


\subsection{Decays of interfaces}
 
A corollary of the
 above calculation  is  a universal formula for the entropy released  in the fusion
 of two conformal interfaces. The result depends only on the angular orientations
of the corresponding branes,   
  \be\label{ggg}
\Delta \log g\,  \equiv\, 
 {\rm log} \left( g( {\cal I}\circ  {\cal I}^\prime)\right)  -  
 {\rm log} \left( g( {\cal I}) \right)
 -  {\rm log} \left( g( {\cal I}^\prime)\right) \nonumber 
 \ee
 \be
 =  {1\over 2}\,  {\rm log} (  1 + {\rm cos}\, 2\vartheta\  {\rm cos}\, 2\vartheta^\prime ) \ . 
  \ee
The sign of the entropy release is the same as the sign of the Casimir
force $d/\varepsilon^2$, where $-d$ is  the leading term in 
the expansion (\ref{euler}).  Both  are
fixed by the product  \   (${\rm cos}\, 2\vartheta\  {\rm cos}\, 2\vartheta^\prime$).   
When this product is  negative  ${\cal I}$  and ${\cal I}^\prime$ tend to attract, 
and their entropy is lowered by fusion. This is in accordance with the prediction of
the g-theorem \cite{gthe}. Conversely, when $\Delta \log g$ is positive 
the composite interface ${\cal I}\circ {\cal I}^\prime$ is an unstable RG  fixed point. 

\vskip 1mm

 One can show that  all non-topological interfaces are
 unstable\footnote{Their fusion with boundaries may, nevertheless,  still
   produce  stable D-branes.}
 by the following argument:   first
strip off their  non-trivial topological part, with the help of the dressing identities
 (\ref{eqdeforme}). What is left behind is the deformed identity operator, separating two regions
 with radii $R\not= R^\prime$.  Notice that these  radii must be different, since otherwise
 the original interface would be topological.  Now the basic radius-jump  interface
  is unstable to splitting into smaller  jumps. Indeed, the dissociation
 \be
 e_{\rm def}^{(R^\prime\leftarrow R)}\  \to \ 
 e_{\rm def}^{(R^\prime\leftarrow R^{\prime\prime})}\circ
  e_{\rm def}^{(R^{\prime\prime}\leftarrow R)} 
 \ee
 is entropically favoured whenever  $R<R^{\prime\prime}<R^\prime$ or  
$R^\prime<R^{\prime\prime}<R$. This follows directly  from  (\ref{ggg}).
 The same conclusion  could in fact  be   reached  by considering
  the effective theory for the radius field,  ${\cal L} \sim (\partial R/R)^2$ ,  
  in which domain walls   tend  to spread to infinite thickness. This  splitting-off of
  radius jumps  tends to push  the bulk radii to their attractor fixed values.  
Conformal interfaces could thus prove to be  a useful tool for studying
  the coupled bulk and boundary  RG flows in string theory. \footnote{Coupled bulk
and boundary RG flows have been studied differently in ref. \cite{keller}.}

  What about the topological interfaces, whose fusion generates no entropy? 
  These are marginally unstable against decay to `prime-factor partons', i.e. $(1,p)$
  or $(p,1)$ interfaces with  $p$ a prime number.  In the case of BPS black holes
  the analogous decays are  hindered by infinite-throats \cite{Seiberg}, so that the
  bound and unbound states can be distinguished. 
  In the case at hand, recombination generally increases the dimension
  of the open-string moduli space and it is unclear whether such a distinction 
  makes sense.    Notice that there is no process which can reduce the entropy
  of the $(1,p)$ ``partons''  back to zero. Annihilation with the  
``antiparton''  $(p,1)$
  releases an entropy  log\,$p$.


   \section{Quantum interfaces}
 
   The interfaces discussed so far connect  two  points in the $S^1$ moduli-space of
 the $c=1$ models. For more general interfaces,
  the  conformal  theories on the two  sides  
   live in different branches of moduli space, or may even be
   completely different theories.
   In the latter case,   it has been  shown by Quella et al \cite{Quella2}
   that the difference of the two central charges,  $\vert c_1 - c_2 \vert $,  
   provides a lower bound
   to the reflectivity of the interface.  Such interfaces are thus never  topological,
   and may be unstable against dissociation processes like those discussed
   in the previous section. This is 
  an interesting question  that  we  will not address here.    
   
\vskip 1mm
 
   Let us consider instead the topological interfaces that connect  the circle
 with  the orbifold branch.  Examples of such interfaces are easy to construct.
  They  include all  D1- branes on $S^1\times
(S^1/Z_2)$ with a $45^o$ orientation. To be more specific, consider 
the interfaces on  the circle line, setting $\alpha =\beta =0$ for
simplicity,
and   with $k_1 = 2l_1$ even. Then
the linear combinations
\be\label{orbcirc}
 (2l_1, \vert k_2 \vert )_{\rm cir/orb} \ \equiv \ 
{1\over 2} (2l_1, k_2)^{(+)} + {1\over 2} (2l_1, -k_2)^{(+)} 
\ee
are good conformal interfaces connecting the circle and the orbifold branch. 
Note that  half-integer coefficients would have been  forbidden for an interface
between circle theories. They are 
here  admissible  because D1-branes and  their images
under $\phi^2$ reflection are
 identified.  As a concrete  example consider the $(2,
1)_{\rm cir/orb}$ interface. It  becomes topological when  the radius of the orbifold 
is double  that of the circle.  Inspection of the lattice sum
(\ref{k1k2i}) shows that this interface projects out the odd-winding sectors
of the circle theory, and the odd-momentum and twisted sectors of the
orbifold theory. It  identifies in an obvious manner the remaining states. 
The entropy of the  map is  ${\rm log} g = {\rm log} \sqrt{2}$. Many other 
circle/orbifold and orbifold/orbifold interfaces can be written down  in a
similar way. 

\vskip 1mm

 These and all previous  topological interfaces  share one important common 
feature: they have a bulk (closed-string) modulus, which
is the product or the ratio of radii on the two sides.
Correspondingly, there is  a semi-classical regime
where their action is, modulo a T-duality transformation, geometrical. 
For the even interfaces, for example, the classical regime is the limit of
large radii with the ratio $R_1/R_2 = \vert k_2/k_1\vert $ \  kept fixed. 
It is well known, on the other hand, that there exist many non-geometric 
D-branes, and the same is true for  conformal interfaces. 
For instance, when  $R_{\rm orb} = 2 R_{\rm cir} = R_*$\  
the orbifold and the circle theories are the same  \cite{Ginsparg},
so  there exists an isomorphism, $\tau$,  between the two.  It is 
certainly  not contained in the list (\ref{orbcirc})  
because it has zero entropy.  When composed with topological maps, from 
 the circle and/or  from the
orbifold side,   it  generates a  whole new class of interfaces,  with 
both the circle and the orbifold  radius  fixed.  We may refer to such non-geometric
interfaces, deep in the CFT moduli space,  as purely `quantum'. 

\vskip 1mm

Quantum interfaces  actually exist also on the circle line. They are
generated by the enhanced  $SU(2)_l\times SU(2)_r$ isometries of the theory at the
self-dual radius, as discussed in ref. \cite{mat2}:
 \be
e^{(R_*\leftarrow R_*)}(h, \tilde h)\ \ \ \ \ {\rm for \ \ all}\ \ \ 
h, \tilde h \in SU(2)\ .  
\ee
Multiplying these isometries with our topological operators,  from left and right,  gives a
large class of topological interfaces : \footnote{Odd interfaces do not give new operators,
because the duality twist is a special  $SU(2)_r$  isomorphism.
Chains of topological operators between two twists also do not produce new operators. 
Such  chains can be always fused  to give an operator with $k_1=k_2$,
 which can then be written as  a superposition
of symmetry generators.} 
 \be
{\cal I}^{(R_1\leftarrow R_2)}(h, \tilde h ;  k_1 , k_2 ;  k_1^\prime , k_2^\prime) 
\ \equiv  \ 
 {\cal I}^{(+)}_{(k_1 , k_2 )} \, \circ\,  e^{(R_*\leftarrow
   R_*)}(h, \tilde h) \, \circ
\,  {\cal I}^{(+)}_{(k_1^\prime , k_2^\prime )}\ .
\ee
For these to be topological both radii must be a priori  fixed,  
\be
R_1 = \left\vert {k_2\over k_1}\right\vert R_* \ \ \ {\rm and}\ \ \
R_2 = \left\vert
{k_1^\prime \over k_2^\prime}\right\vert R_*\  \   . 
\ee
The above operators reduce, in fact, to the even and odd interfaces of the previous
sections when $h$ and $\tilde h$ commute (up to reflection) with the $U(1)^2$ 
generators   of  the circle line. In this special situation, the constraint on
one combination of  radii
gets relaxed. 
For more general rotations, these interfaces break
all   $U(1)$ symmetries of the model.  As shown in ref. 
\cite{mat2} , 
their action on the basic D-branes  of the circle
theory  produces   the continuous extrapolations between arrays of D0
and D1 branes that were  constructed in ref. \cite{GabRec}.  This  shows that
topological defects can act on D-branes in non-trivial ways. 

\vskip 1mm
 
It would be very interesting to extend the analysis of our paper  to these, purely
quantum,  generators of the interface algebra.  This is not straightforward, because
it is unclear how to separate these  topological dresses in a
fusion process.  The results of ref.  \cite{keller} actually 
suggest that the radius deformations  of the
enlarged algebra may be singular.   We hope to return to these
questions in the near future.

\vskip 0.3 cm
{\bf Acknowledgements}
\vskip 2mm
We thank Matthias Gaberdiel, 
Bernard Julia, Daniel Roggenkamp and Jan Troost for discussions.
This work was supported by a EURYI award of the European
Science Foundation (I.B.), and by the European Networks 'Superstring Theory'
(MRTN-CT-2004-512194) and  `Forces Universe' (MRTN-CT-2004-005104).


\end{document}